\documentclass{ADAPTCONF2025}
\usepackage{balance}
\usepackage{listings}            % For code listings
\usepackage{xcolor}              % Custom colors in listings
\usepackage{colortbl}
\usepackage{tcolorbox}
\tcbset{
  colback=white,
  colframe=blue!40!black,
  coltitle=blue!20!black,
  fonttitle=\bfseries,
  boxsep=2pt,
  arc=3pt
}

\definecolor{headerblue}{RGB}{220,230,255}
\definecolor{rowgray}{RGB}{245,245,245}

\usepackage{tikz}                % For diagrams
\usetikzlibrary{positioning, arrows.meta}  % TikZ libraries
\usepackage{adjustbox}           % Resize tables or figures

\interspeechcameraready

\title{Learning-Infused Formal Reasoning: Contract Synthesis, Artifact Reuse and Semantic Foundations}

\name{Arshad Beg$^1$, Diarmuid O'Donoghue$^1$, Rosemary Monahan$^1$}

\address{
$^1$Department of Computer Science, Maynooth University, Ireland
}

\email{\{arshad.beg, diarmuid.odonoghue, rosemary.monahan\}@adaptcentre.ie}

\begin{document}

\maketitle

\begin{abstract}
Artificial intelligence systems have achieved remarkable capability in natural language processing, perception and decision-making tasks. However, their behaviour often remains opaque and difficult to verify, limiting their applicability in safety-critical systems. Formal methods provide mathematically rigorous mechanisms for specifying and verifying system behaviour, yet the creation and maintenance of formal specifications remains labour intensive and difficult to scale. This paper outlines a research vision called \textit{Learning-Infused Formal Reasoning (LIFR)}, which integrates machine learning techniques with formal verification workflows. The framework focuses on three complementary research directions: automated contract synthesis from natural language requirements, semantic reuse of verification artifacts using graph matching and learning-based embeddings, and mathematically grounded semantic foundations based on the Unifying Theories of Programming (UTP) and the Theory of Institutions. Together these research threads aim to transform verification from isolated correctness proofs into a cumulative knowledge-driven process where specifications, contracts and proofs can be synthesised, aligned and reused across systems.
\end{abstract}

\section{Introduction}

This paper is a short version of our accepted paper \cite{Beg2026LIFR}.
\vspace{0.3cm}

Artificial intelligence has rapidly transformed many areas of computing,
enabling systems capable of interpreting natural language, analysing complex
data and supporting complex decision-making processes. Large language models
and other learning-based systems have demonstrated remarkable capability in
tasks that traditionally required human reasoning, including code generation,
knowledge extraction and automated analysis of textual artefacts. Despite
these advances, such systems often behave as opaque models whose internal
reasoning is difficult to interpret or verify. This lack of transparency
creates significant challenges in safety-critical domains such as
cyber-physical systems, autonomous platforms and high-assurance software
systems, where correctness and reliability must be formally guaranteed.

Formal methods provide a complementary paradigm for addressing these
challenges. By expressing system behaviour using mathematically precise
specifications and verification techniques such as model checking and theorem
proving, formal methods enable developers to reason rigorously about program
correctness. Over several decades, these techniques have been successfully
applied in domains including hardware verification, safety-critical software
and protocol analysis. However, the widespread adoption of formal methods
remains limited by several practical barriers. Constructing formal
specifications requires specialised expertise, translating informal
requirements into logical constraints is labour intensive, and verification
artifacts such as specifications and proofs are rarely reused across projects.

At the same time, recent advances in machine learning offer new opportunities
to address these limitations. Learning models are particularly effective at
recognising patterns in large datasets and extracting structured information
from natural language descriptions. When combined with symbolic reasoning
systems, these capabilities create the possibility of \emph{neuro-symbolic}
verification workflows in which learning techniques assist with the discovery
and synthesis of formal artifacts while logical reasoning ensures their
correctness. Rather than replacing formal methods, machine learning can act
as an enabling technology that reduces the cost of specification development
and increases the reuse of existing verification knowledge.

This paper outlines a research vision called \textit{Learning-Infused Formal
Reasoning (LIFR)}, which explores the integration of learning-based methods
with formally grounded verification frameworks. The goal of LIFR is to
transform formal verification from an isolated activity focused on individual
program proofs into a cumulative knowledge-driven process in which
specifications, contracts and proofs can be synthesised, discovered and
reused across systems. The framework focuses on three complementary research
directions. The first investigates automated contract synthesis from natural
language requirements using learning-assisted specification generation. The
second explores graph-based representations and matching techniques to enable
semantic reuse of verification artifacts across projects. The third develops
rigorous semantic foundations based on the Unifying Theories of Programming \cite{hoare1998unifying}
and the Theory of Institutions \cite{Tarlecki2005Institutions}, providing the mathematical infrastructure
necessary to support heterogeneous verification environments.

Together these research threads aim to create an ecosystem in which learning
techniques assist with the discovery and reuse of formal knowledge while
strong semantic foundations ensure correctness and interoperability. The
following sections describe these research directions and outline how they
collectively contribute to the long-term vision of learning-infused formal
reasoning.

\section{Contract Synthesis}

A persistent challenge in formal verification is the translation of informal
requirements into precise, machine-verifiable specifications. Software
requirements are typically written in natural language and therefore contain
ambiguities, implicit assumptions and contextual dependencies that make direct
formalisation difficult. Program contracts—such as preconditions,
postconditions and invariants—provide a structured mechanism for expressing
expected behaviour, yet they are often written manually and only after
significant development effort. This manual process limits scalability and
creates a bottleneck for the adoption of formal methods in industrial
settings.

Our research vision therefore explores automated contract synthesis within
the broader LIFR framework. Inspired by the VERIFYAI architecture (Figure 1b of \cite{BegEtAl2025}) 
proposed in our earlier work \cite{BegEtAl2025}, the approach treats specification generation as a
neuro-symbolic pipeline. Large language models interpret natural language
requirements and generate candidate logical predicates capturing expected
behavioural constraints. These candidates are then evaluated using formal
verification engines such as SMT solvers \cite{robere2018proof} or theorem provers to ensure
satisfiability and logical consistency. Iterative feedback from the
verification stage guides refinement of generated contracts, enabling the
system to progressively converge toward valid and verifiable annotations.

Beyond simple translation, the long-term objective is to enable contracts to
become dynamic artifacts that evolve alongside software systems. Rather than
remaining static documentation, contracts could be inferred, repaired and
extended automatically as code evolves or as new requirements emerge. In such
a setting, learning-based synthesis would operate within a formally grounded
verification loop, ensuring that generated specifications remain faithful to
their semantic intent while benefiting from the pattern-recognition
capabilities of modern AI models.

\section{Graph Matching for Artifact Reuse}

While generating new specifications is important, the verification community
already possesses a large body of existing knowledge in the form of verified
specifications, proof scripts and modelling artifacts. Unfortunately, these
resources are rarely reused systematically because they are expressed in
different syntactic forms and often belong to different modelling frameworks.
The LIFR agenda therefore explores mechanisms for semantic reuse of
verification artifacts through hybrid graph-based reasoning combined with
learning-based semantic analysis.

In this approach, programs, specifications and proofs are represented as typed
and attributed graphs that capture structural relationships between program
elements, logical predicates and verification obligations. Graph
representations provide a natural way to model dependencies among variables,
constraints and behavioural properties. Once artifacts are represented as
graphs, approximate graph matching algorithms can be used to identify
structural similarities between existing verification artifacts and new
verification tasks. Figure 2 of \cite{Beg2026LIFR} depicts the workflow
of the process. 

To complement structural similarity, large language models are used to enrich
graph nodes with semantic embeddings derived from identifier names,
documentation and comments. This hybrid representation allows matching
algorithms to combine structural correspondence with semantic similarity.
When potential matches are identified, graph transformation rules can adapt
retrieved artifacts to fit new contexts. These transformations may involve
refining predicates, adjusting abstraction levels or restructuring proof
components. Over time, such mechanisms could enable verification workflows to
benefit from cumulative knowledge, where previously verified artifacts
accelerate the analysis of new systems.

\section{Semantic Foundations using UTP}

Although learning techniques can assist with synthesis and reuse, trustworthy
verification requires strong mathematical foundations. The LIFR research
programme therefore emphasises semantic frameworks capable of supporting
heterogeneous modelling languages and verification tools. Two complementary
theoretical frameworks provide the foundation for this effort: the Unifying
Theories of Programming (UTP) and the Theory of Institutions.

UTP provides a relational semantic model that unifies diverse programming
paradigms through refinement relations and healthiness conditions. Within
UTP, imperative programs, concurrent processes and state-based specifications
can be interpreted within a single algebraic framework. This unified view
allows reasoning principles such as refinement, composition and behavioural
equivalence to be applied consistently across different modelling paradigms.
Recent advances such as Isabelle/UTP have further demonstrated the potential
for mechanised reasoning within this semantic framework.

Complementing UTP, the Theory of Institutions provides an abstract,
category-theoretic framework for representing logical systems through
signatures, sentences, models and satisfaction relations. Institutions enable
formal reasoning about relationships between different specification
languages and support satisfaction-preserving translations across modelling
frameworks. Together, UTP and institutional semantics provide a rigorous
foundation for interoperability across verification environments.

Within the LIFR vision, these theories act as a form of ``semantic governance''
for AI-assisted verification workflows. Learning systems may generate
candidate specifications or transformations, but these artifacts are
interpreted and validated within the constraints imposed by UTP healthiness
conditions and institutional semantics. This ensures that AI-generated
artifacts remain logically sound while enabling interoperability between
different verification tools and modelling languages.

\section{Conclusion}

This paper outlines a research vision for LIFR, a framework that integrates machine learning with formal verification to support scalable and reusable specification workflows. The proposed agenda combines automated contract synthesis, graph-based artifact reuse and rigorous semantic foundations grounded in UTP and institutional logic.

By combining learning-driven discovery with symbolic reasoning guarantees, the LIFR framework aims to transform verification into a cumulative knowledge-driven process. In such an ecosystem, specifications and proofs can be systematically synthesised, aligned and reused across systems, enabling more scalable and trustworthy engineering of complex and AI-enabled software systems. 

\section{Acknowledgements}
This work is partly funded by the ADAPT Research Centre for AI-Driven Digital Content Technology, which is funded by Research Ireland through the Research Ireland Centres Programme and is co funded under the European Regional Development Fund (ERDF) through Grant 13/RC/2106 P2. The submission aligns with Digital Content Transformation (DCT) thread of the ADAPT research centre.
\balance

\bibliographystyle{IEEEtran}
\bibliography{AASC2026Bib}

\end{document}